# Growth of large-sized relaxor ferroelectric PZN-PT single crystals by modified flux growth method

P. Vijayakumar[a], C. Manikandan[b], R.M. Sarguna[a], Edward Prabu Amaladass[a,c], K. Ganesan[a,c], Varsha Roy[a], E. Varadarajan[b,d], S. Ganesamoorthy[a,c,*]

[a] *Materials Science Group, Indira Gandhi Centre for Atomic Research, Kalpakkam, 603 102, India*
[b] *DRDO Industry Academia (DIA)-Ramanujan Centre of Excellence (RCoE), IITM Research Park, Chennai-600113, India*
[c] *Homi Bhabha National Institute, Training School Complex, Anushakti Nagar, Mumbai, 400094, India*
[d] *Combat Vehicle Research and Development Establishment (CVRDE), Avadi, Chennai-600054, India*

*Email: sgm@igcar.gov.in

## Abstract

A novel bottom-cooling high-temperature solution growth technique is developed for growing large-sized relaxor ferroelectric $0.91Pb(Zn_{1/3}Nb_{2/3}O_3)$-$0.09PbTiO_3$ (PZN-PT) single crystals. During the growth, an inverse temperature gradient is maintained in the crucible base by flowing air at a controlled rate. This method restricts the number of spontaneously nucleated crystals at crucible bottom, reduces loss of volatile PbO component and favours the growth of large-sized PZN-PT single crystals. Large-sized PZN-PT single crystals of dimensions ~ 22×20×14 mm$^3$ are reproducibly grown by the proposed method. The electrical characteristics of the PZN-PT wafers oriented along the <100>, <010> and <001> directions are investigated. PZN-PT wafers oriented along the <001> direction exhibited superior piezoelectric coefficient ($d_{33}$) of ~ 2221 pm/V. The homogeneity of the physical parameters is analysed by preparing 10 elements with dimensions of ~5×2.5×2.5 mm$^3$ which were cut from single wafer oriented along the <001> direction. The ferro-, piezo- and dielectric characteristics of these wafers were found to be highly uniform with small standard deviation. The observation of $d_{33}$ value with less than 2 % deviation from mean value confirms the growth of high quality PZN-PT single crystals.

1. Introduction

The relaxor ferroelectric single crystals of (1-$x$)Pb(Zn$_{1/3}$Nb$_{2/3}$)O$_3$–$x$PbTiO$_3$ ($0 \leq x \leq 0.1$) near the morphotropic phase boundary (MPB) between its rhombohedral and tetragonal phase have been studied extensively [1–23]. Single crystal of 0.91Pb(Zn$_{1/3}$Nb$_{2/3}$)O$_3$-0.09PbTiO$_3$ (PZN-PT), show anomalously large electro-mechanical coupling factor ($k_{33} > 0.9$); high piezoelectric coefficient ($d_{33} > 2000$ pm/V); low coercive field ($E_c < 5$ kV/cm) and a high dielectric constant ($\varepsilon_r > 5000$), in comparison to established commercial lead zirconium titanate (PZT) ceramics whose $d_{33}$, $E_c$, and $\varepsilon_r$ values are about 700 pm/V, 30 kV/cm and 4000, respectively [15,16,18]. Young's modulus of PZN-PT is 70 - 80 % lower than PZT and thereby permitting smaller transducer size for a vital frequency range [10]. During the last two decades, research and development have been focused on to improve the sensitivity and bandwidth of ultrasonic transducers using PZN-PT single crystal instead of the conventional PZT transducers. PZN-PT is proposed to be the best candidate for high-performance piezo-devices such as medical ultrasound imaging probes, sonars for underwater communications, high-authority sensors and actuators [3,9,11,20–22].

Wide spread deployment of PZN-PT transducers has been hindered owing to the limited availability of high-quality large-sized single crystals. The incongruent melting of the PZN-PT compound restricts the growth of crystals from stoichiometric melts. Myl'nikova and Bokov [24] pioneered in the growth of relaxor ferroelectric single crystals and grown Pb(Zn$_{1/3}$Nb$_{2/3}$)O$_3$ (PZN) crystals by solution growth technique using PbO as flux. Nomura and co-workers [4] grew successfully single crystals of (1-$x$)PZN-$x$PT solid solutions in the entire composition range. Later, several methods had been adopted for the growth of PZN-PT single crystals from fluxes [5,7,16,17,19,25–29]. Kobayashi et.al. [30] grew large-sized 0.91PZN-0.09PT single crystal with flux to solute ratio varying from 70 : 30 to 50:50 at cooling rate of ~ 1 °C/h at saturation temperature of 1260 °C using modified flux method with bottom cooling by oxygen flow upto 1 l/min. Similarly, Kumar et. al. [26], and Lim and Rajan [19] grew large-sized 0.91PZN-0.09PT single crystal with flux to solute ratio of 45: 55 at cooling rate of ~ 1 °C/h down to 900 °C starting from the saturation temperature of 1200 °C using bottom-cooled flux method with oxygen flow at the bottom of the Pt crucible. Xu et.al. [31] grew PZN-9%PT single crystal using 50:50 solute to flux ratio, saturation temperature 1200 °C with dwelling time of 6 h, translation rate at 0.5 mm/h upto 900 °C and gas flow rate of 1.6 l/min by solution Bridgman methods. Dabkowski et.al. [6] modified the bottom cooling arrangement by point bottom cooling with Pt cold finger arrangement for successful removal of latent heat formation during crystal growth. In our earlier study we grew the PZN-PT crystal using modified flux method with flux to solute ratio of 60:40 [21].

PZN-PT single crystals are an indispensable for the development of high end technology related to transducer applications. Although several reports exist in the literature for the growth of large-sized high-quality PZN-PT single crystals using high-temperature solution growth, it is not trivial to grow large size crystal due to many technological challenges involved in the crystal growth. This motivates us to develop in-house modified flux growth method for of growth of large-sized PZN-PT single crystals which are essential for developing transducer devices.

## 2. Experimental

In-house designed crystal growth system with necessary heating module, bottom cooling arrangement and provision for decanting of the crucible after growth were developed for growing PZN-PT single crystals. Figure 1a shows the schematic of our modified flux growth system and the important components are listed in the figure caption. The complete growth system was enclosed in a leak proof enclosure. High purity (~ 99.95%) starting materials such as PbO, ZnO, $Nb_2O_5$ and $TiO_2$ were taken in the ratio of 55 mol % PbO: 45 mol % 0.91PZN-0.09PT. The growth runs were performed for 300 g inclusive of the solvent weight. The materials were then mixed in a three-dimensional shaker mixer and loaded into a platinum crucible. The platinum crucible was then closed with a platinum lid and then placed in an alumina crucible. The free space inside the alumina crucible was filled with alumina powder and then sealed using alumina lid by ceramic binder. This strategy limits PbO weight loss to less than 2 % in an overall growth period of 600 h. The loaded crucible was then placed in a suitable temperature region of developed resistive heated furnace. The furnace temperature was increased to 1250 °C and soaked for ~ 5 - 10 h and then the temperature was lowered down to the saturation temperature of 1150 °C. In order to reduce the number of nuclei and also to induce growth at the crucible base centre, a provision for air cooling through a fine nozzle to the crucible base is provided. The flow rate of air was varied in the range of 0.5 - 2 l/min. By this arrangement a large inverse temperature gradient (~ 50 - 75 °C) between crucible base and top was imparted such that growth was initiated at the base of the crucible. Then growth was commenced with progressive cooling rate of ~ 0.5 - 2.0 °C/h down to 900 °C. At this temperature, the growth crucible was decanted and cooled at a faster rate (50 °C/h) to room temperature.

Growth optimization was performed by varying the air flow rate and cooling protocol. Figure 1b-e shows the crystals that are grown under different air flow rate and solution cooling rate. When both air flow rate and solution cooling rate were high ~ 2 l/min and 2 ºC/hr respectively, multiple small sized crystals (Fig 1b) were grown. On further reduction in the cooling rate (~ 2 l/min and 1 ºC/h) the crystal became poly-crystalline embedded with large grain size crystals (Figure 1c). On further reduction in air flow rate (1 l/min and 0.5 l/min) and cooling program (0.5 ºC/h) resulted in large grain samples as depicted in Figure 1d and Figure 1e, respectively. We note here that when both the air flow rate and the solution cooling rate were high, or either one of them was high, the degree of super-saturation of the solution becomes high at the specified location at the bottom of the crucible and thus forms multi-nucleation [14]. After completion of the growth run, the crucible was removed from the furnace. The platinum crucible was leached with hot $HNO_3$ to remove the PZN-PT crystal from PbO flux.

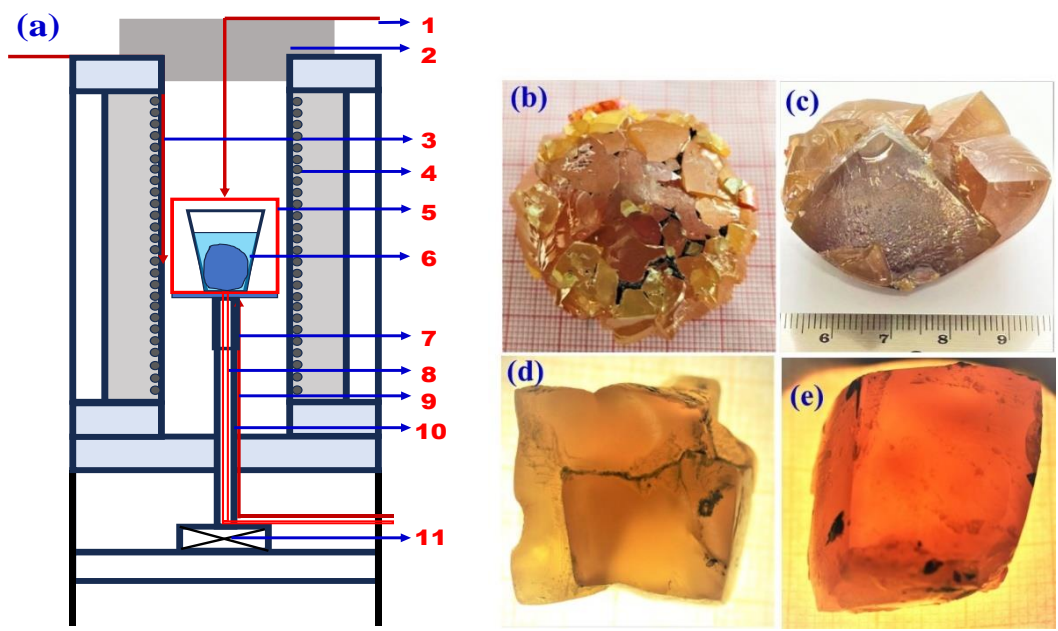

Figure 1. (a) Schematic of the modified flux growth system: (1) top thermocouple, (2) top insulation, (3) controlling thermocouple, (4) heater, (5) alumina crucible, (6) platinum crucible with PZN-PT material, (7) platinum foil, (8) gas tube, (9) bottom thermocouple, (10) alumina tube and (11) adjustable stand. (b-e) The PZN-PT crystals grown at different air flow rate and solution cooling rate, (b) 2.0 l/min & 2 ºC/h, (c) 2.0 l/min & 1 ºC/h, (d) 1.0 l/min & 0.5 -1 ºC/h and (e) 0.5 l/min & 0.5-1.0 ºC/h

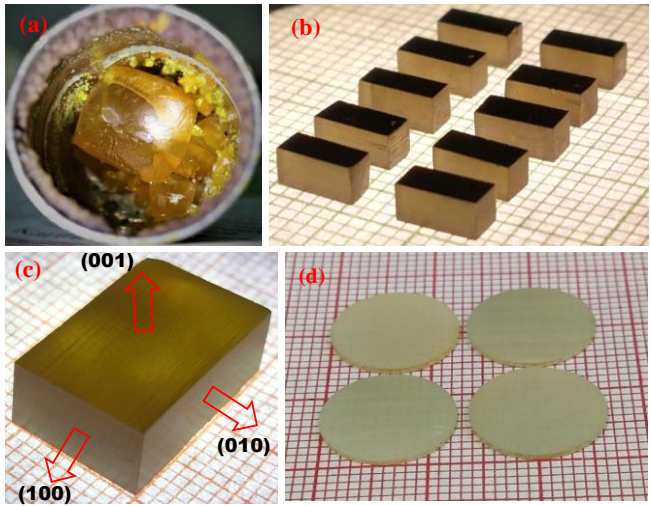

Figure 2: A photograph of (a) as-grown PZN-PT single crystal inside platinum crucible, (b-d) oriented transducer elements with various shapes and dimensions.

Figure 2a shows large-sized as-grown PZN-PT single crystal inside platinum crucible and the dimension of the grown crystal is ~ 22×20×14 mm$^3$. This crystal is grown at the optimized growth parameters of the temperature gradient of ~ 50 - 100 °C between top and bottom of the crucible, air flow rate of 0.5 l/min and the solution cooling rate of 0.5-1 °C/hr down to 900 °C. The grown crystal is translucent and light brownish-yellow in colour with natural habitual growth faces. The weight of each grown crystal from the successive growth runs conducted was in the range of 60 - 100 g depending upon the growth parameters for the initial charge of 300 g. The grown PZN-PT single crystal was mounted on 5-axis goniometer and oriented along the <100>, <010> and <001> crystallographic direction using back scattered X-ray Laue diffraction method. The diffraction patterns were simulated using ORIENTX software and compared with recorded diffraction pattern. Crystal was then diced along crystallographic planes using desktop crystal cutter. Then, PZN-PT single crystal elements were diced into different shapes like rectangular bar and circular disc elements for various device applications, as shown in Figure 2b-d. To study the homogeneity of the crystal, a large wafer with dimension of ~ 15×11×2.5 mm$^3$ was cut along the <001> orientation and then the wafer was further diced into 10 elements with dimensions of ~ 5×2.5×2.5 mm$^3$. The diced wafers were lapped using silicon carbide sheets of grit size 800, 2000, and 4000 respectively, to get smooth surface. The final polishing was done with Al$_2$O$_3$ powders of 0.5 and 0.05 μm.

The wafers were then annealed at 200 °C for 3 h to remove the strain formed during element preparation process such as dicing, lapping and polishing. For electrical contacts,

Au/Cr was sputtered on the both face of PZN-PT single crystal element by DC magnetron sputtering (Moorfield, Minilab 060, Manchester, U.K.). On the cleaned element surface, Cr (30 nm) was deposited as adhesion layer and followed by Au (300 nm) layer at room temperature. The ferroelectric and piezoelectric properties of PZN-PT elements were measured by TF Analyzer 2000 (aixACCT Systems GmbH, Aachen, Germany). The element was poled near Curie temperature. During application of electric field the crystal was immersed in silicone oil bath to avoid electric arcing. Bipolar voltage in a sinusoidal form with 1 Hz frequency is applied using high voltage amplifier (10/10B-HS, Trek, Inc. New york, USA). The applied voltage is gradually increased with step size of 100 V starting from 100 V to 4 kV. Electro-mechanical characteristics as a function of frequency (40 Hz to 10 MHz) were measured using precision impedance analyzer (Agilent 4294A, California, USA). Dielectric characteristics of PZN-PT single crystal are measured using Impedance Analyzer (E4990A Keysight, Santa Rosa CA, USA). The unipolar strain measurements were carried out using aixACCT Systems GmbH, Aachen, Germany, on the basis of interferometric technique. The dielectric constant and dielectric loss were studied from room temperature to 220 C using a LCR meter (HIOKI 3532-50, Japan).

## 3. Results and discussion

### 3.1 X-ray diffraction

The prepared <001>, <010> and <001> orientation of the elements were confirmed by Laue diffraction pattern. The periodic arrangements of the bright diffraction pattern confirms the single crystal and oriented along the <001> direction as shown in Figure 3a. A small portion of the grown crystal was ground into fine powders and the powder XRD pattern is shown in Figure 3b. It confirms that the grown crystals have pervoskite structure with tetragonal structure and the lattice parameter values are found to be $a = b = 4.049$ Å and $c = 4.056$ Å and the values are consistent with the literature [32].

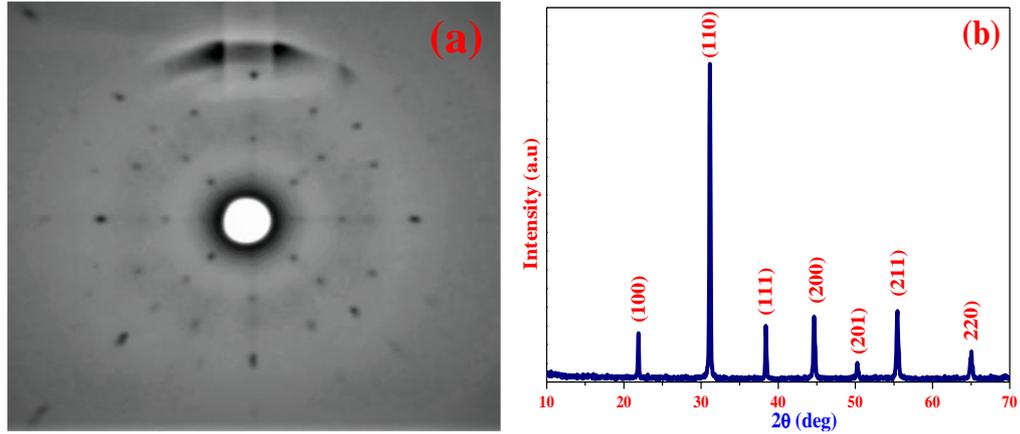

Figure 3: (a) Laue diffraction pattern on the <001> wafer and (b) powder X-ray diffraction pattern of PZN-PT

## 3.2. Ferroelectric, piezoelectric and strain analysis

Ferroelectric hysteresis response to applied electric field along different (100), (010) and (001) crystallographic planes were measured and it is shown in Figure 4a. PZN-PT single crystals exhibit a well saturated hysteresis loop. For an applied frequency of 1 Hz, the maximum electric field is about 5.4, 5.4 and 4.9 kV/cm along (100), (010) and (001) planes, respectively. The saturation polarization is about 24.9, 27.5 and 29.5 $\mu C/cm^2$ along (100), (010) and (001), respectively. The strain measurement is recorded at room temperature after poling for 24 h.

The unipolar strain is measured on the basis of interferometric technique wherein the monochromatic He-Ne laser (632 nm) beam is made to fall on top side of the crystal. The sinusoidal form of applied voltage is increased gradually and corresponding strain response was measured. As shown in Figure 4b, the unipolar strain-electric field curve displays nearly a hysteresis free behavior as an outcome of stable domain configuration in the poled direction and the strain value upsurges linearly with electric field till 5 kV/cm. From the strain versus electric field curves, it is found that the strain values were zero after removal of electric field, i.e., $\Delta S_{E=0} = 0$ which implies no strain remanence effect after poling in the PZN-PT crystals. The measured strain values are 0.04, 0.12 and 0.16 %, and the estimated longitudinal $d_{33}$ values are 1195, 1970 and 2221 pm/V for the <100>, <010> and <001> oriented wafers, respectively. The area of the strain-field response curve specifies that the PZN-PT samples are in the vicinity of MPB composition. These measurements confirm that the <001> wafer possesses higher saturation polarization, remnant polarization, strain, piezocoefficient and low coercive field as compared to that of the <100> and <010> orientations. We note here that the piezoelectric charge coefficient was also measured by PM300, Piezotest Pte. Ltd., Singapore which is based on the quasistatic or 'Berlin court' method. The measured piezoelectric co-efficient obtained by the unipolar strain method was found to have the same value.

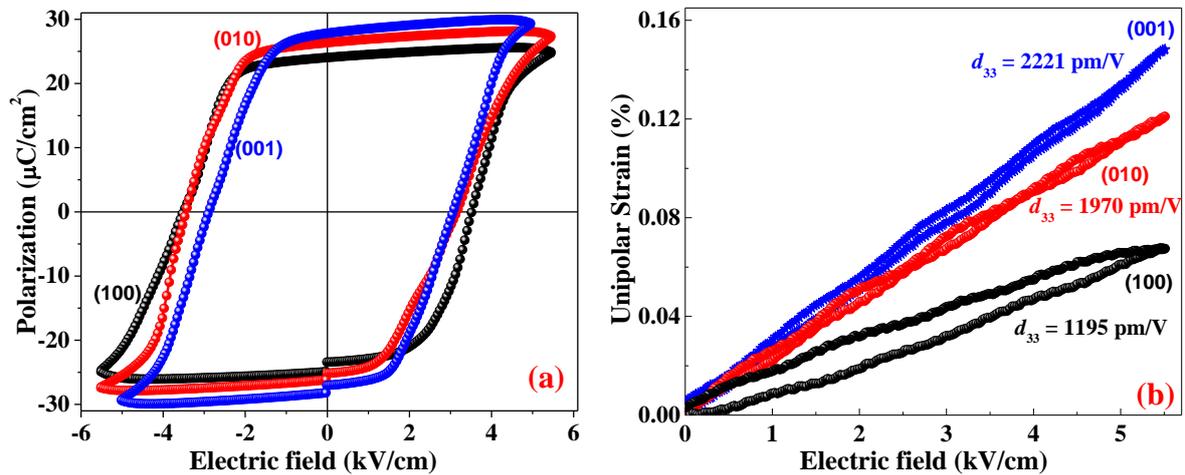

Figure 4: (a) P-E loop and (b) unidirectional strain loop for the (100), (010) and (001) crystallographic planes of PZN-PT

### 3.3. Impedance analysis

The impedance spectra of the poled and oriented PZN-PT wafers together with phase angle variation in response to applied electric field at varied frequencies (100 Hz to 2 MHz) were measured at room temperature and the results are shown in Figure 5a-c. The dimension of the PZN-PT wafers used for impedance spectra is 5×5×2 mm$^3$ which resembles square plates. The thickness mode of resonance is used in the impedance spectra measurements. The measured fundamental resonant frequency ($f_r$) at minimum impedance is 1.35, 2.02, and 2.19 MHz and the anti-resonant frequency ($f_a$) at maximum impedance is 1.46, 2.35, and 2.29 MHz along the (100), (010) and (001) planes, respectively. The magnitude of impedance is of the order of 1 kΩ indicating the resistive nature of the sample even after the poling process. The conductance and susceptance with frequency were measured along the oriented planes and are shown in Figure 5d-f. Conductance - susceptance curve displays typical ferroelectric nature as an inversion of impedance - phase angle curve behavior. PZN-PT single crystals are prone to chipping on the edges due to its lower hardness that causes multi-peaks in the impedance spectra, particularly more visible near the resonance frequency. In addition, the impedance spectra of standard piezoelectric vibrator of the (001) PZN-PT wafers with rectangular bar (5.5×2.5×2.5 mm³) and circular disc (10 mm dia and 1 mm thick) geometries were also measured and displayed typical behavior as similar to Figure 5c at different frequencies.

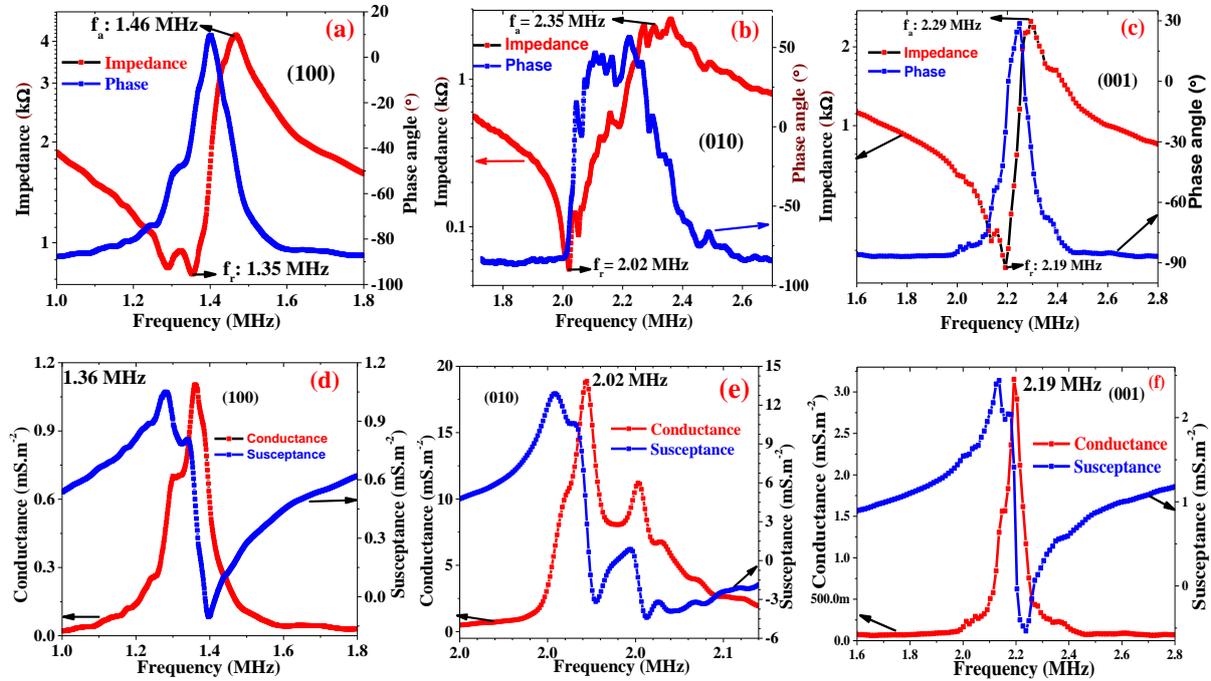

Fig. 5: (a-c) The variation of impedance and phase and (d-f) conductance and susceptance as a function of frequency of PZN-PT wafers along the <100> (a,d), <010> (b,e) and <001> orientations (c,f).

We also recorded the impedance and phase angle as a function of frequency and temperature dependent dielectric constant and dielectric loss for the PZN-PT wafers oriented along the (001) direction at 100 kHz and the results are shown in Figure 6a and 6b, respectively. Here the dielectric constant exhibits two peaks around 95 °C and 184 °C that correspond to rhombohedral to tetragonal phase and tetragonal to cubic phase transition temperatures, respectively. The relaxor phenomenon was confirmed by the observation of diffused phase transition around 184 °C. The measured maximum value of dielectric constant and dielectric loss are ~ 14287 and 0.376 respectively, at Curie temperature.

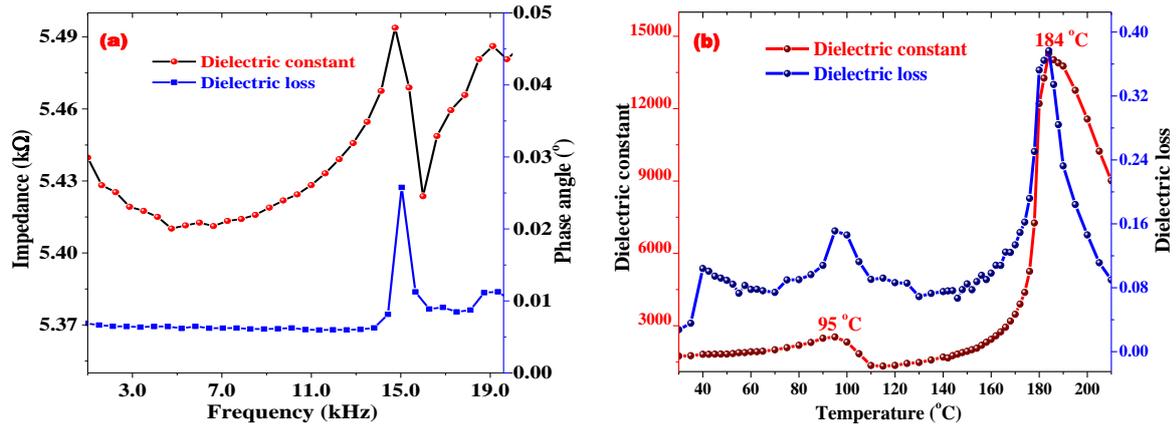

Figure 6. (a) Dielectric constant and dielectric loss spectra as a function of frequency at room temperature, (b) temperature dependence of dielectric constant and dielectric loss of PZN-PT crystal along the crystallographic plane of (001).

### 3.4. Crystal homogeneity analysis

The grown PZN-PT single crystal's structural and compositional uniformity was analyzed by ferro, piezo and dielectric measurements on 10 elements with dimensions of $5 \times 2.5 \times 2.5$ mm$^3$ that were diced in a single (001) wafer across the whole crystal. The diced PZN-PT elements are shown in Fig 2b. The Cr/Au electrode was deposited on the (001) plane and they were poled at 3.8 kV/cm field at room temperature. The ferroelectric, piezoelectric, and dielectric behavior of these 10 elements was analyzed. Figure 7a depicts the P-E loop measured over the 10 elements and these loops appear nearly identical. Similarly, the other experimental data such as unidirectional strain and impedance spectroscopy display a nearly identical behavior. The variation of the extracted parameters such as $d_{33}$ and dielectric constant, $k_{33}$, and piezoelectric voltage coefficient ($g_{33}$) and resonance and anti-resonance frequency for the 10 elements are shown in the Figure 7b, 7c and 7d, respectively. The values of the dielectric and piezoelectric parameters are comparable to the values reported in the literature [19,32]. Table 1 provides the statistical value of the parameters viz. remnant polarization, saturation polarization, dielectric loss, resonance and anti-resonance frequency, strain, $d_{33}$, $k_{33}$ and $g_{33}$ values which were estimated from the above measurements. As can be seen from Table 1, the statistical error values of the physical parameters among the 10 elements are very minimal. It should be noted here that the standard deviation of $d_{33}$ value is only about 2 % from the mean value across the whole crystal. Such a small deviation in the physical parameters ensures the high structural homogeneity throughout the single crystals grown by the improved flux growth technique.

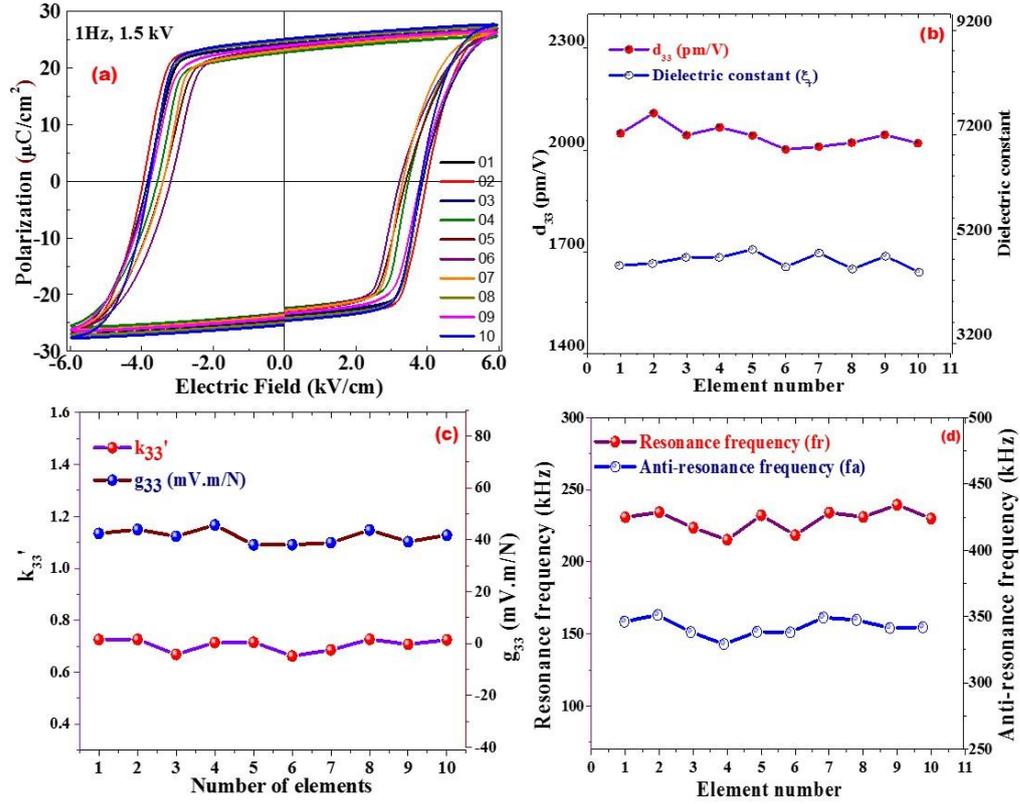

Figure 7. Measured properties (a) P-E loop, (b) $d_{33}$ and dielectric constant, (c) $k_{33}'$ and $g_{33}$, and (d) $f_r$ and $f_a$ versus number of developed element.

**Table. 1**. **Comparison of ferroelectric, piezoelectric and dielectric properties of 10 numbers of the <001> oriented PZN-PT single crystal wafers with literature data**

| Materials Parameter | Parameter range (Min – Max) | Statistical mean | Reported values [32] |
|---|---|---|---|
| Remnant polarization $P_r$ μC/cm$^2$ | 23.0– 25.2 | 24.1±0.7 | 23 – 30 |
| Sat. polarization $P_{max}$ μC/cm$^2$ | 25.55 – 27.62 | 26.6±0.62 | 24 – 32 |
| Coercive field $E_c$ kV/cm | 3.61 – 3.95 | 3.72±0.09 | 3.2 – 4.2 |
| Dielectric loss (tan δ) % @ 1kHz | 0.007–0.028 | 0.02±0.006 | 0.11 – 0.9 |
| Dielectric constant $\varepsilon_r$ | 5160-5934 | 5552±311 | 4500 – 6000 |
| Resonance frequency $f_r$ kHz | 215.2-239.5 | 229.0±7.6 | - |
| Anti resonance frequency $f_a$ kHz | 291.3-351.2 | 338.4±17.2 | - |
| Piezoelectric coefficient $d_{33}$ pm/V | 1962-2086 | 2024±39 | 1800 – 2700 |
| Piezoelectric voltage coefficient $g_{33}$ mV. m/N | 38.03-45.67 | 41.3±2.6 | 45.5 – 50.43 |
| Strain % | 0.1055–0.1369 | 0.1206±0.0098 | 0.14 – 0.15 |

## 4. Conclusions

Large-sized high-quality relaxor ferroelectric $0.91Pb(Zn_{1/3}Nb_{2/3}O_3)$-$0.09PbTiO_3$ (PZN-PT) single crystals are successfully grown by modified flux growth method. Among the <100>, <010> and <001> oriented wafers, the <001> oriented element possesses higher remnant polarization $P_r$ of 24.82 μC/cm$^2$, and the maximum unipolar strain of 0.16 % with zero remnant strain. The electrical characteristics of the grown PZN-PT crystals is found to be extremely uniform throughout the crystal as evidenced by the ferro-, piezo- and dielectric measurements. The deviation of the piezoelectric coefficient $d_{33}$ value is less than 2 % from the mean value across the crystal asserting the uniformity of the grown PZN-PT single crystals. Thus, the modified high temperature solution growth method is one of the most suitable techniques for the growth of various relaxor ferroelectric single crystals having large size with structural homogeneity.

## 5. Acknowledgements

The authors acknowledge the support and encouragement from Dr. B.Venkatraman, Director, IGCAR. Thanks are also to Dr. R. Divakar, Group Director, MSG and MMG and R.Nagendran, Head, SDTD for their support and encouragement. The authors thank Dr.V.Natarajan, DRDO and Prof. M.S.Ramachandra Rao, IITM, Chennai for support and encouragement. P.Vijayakumar is thankful to IGCAR for the award of Research Associate and Visiting Scientist fellowships.

### References

[1] H. Zhou, T. Li, N. Zhang, M. Mai, M. Ye, P. Lin, C. Huang, X. Zeng, H. Huang, S. Ke, A diagram of the structure evolution of $Pb(Zn_{1/3}Nb_{2/3})O_3$-9%$PbTiO_3$ relaxor ferroelectric crystals with excellent piezoelectric properties, Crystals. 7 (2017) 130. doi:10.3390/cryst7050130.

[2] E. Sun, W. Cao, Relaxor-based ferroelectric single crystals: Growth, domain engineering, characterization and applications, Prog. Mater. Sci. 65 (2014) 124–210. doi:10.1016/j.pmatsci.2014.03.006.

[3] J. Kim, Y. Roh, Equivalent properties of 1-3 piezocomposites made of PMN-PT single

crystals for underwater sonar transducers, Behav. Mech. Multifunct. Mater. Compos. 7978 (2011) 797826. doi:10.1117/12.879610.

[4]   S. Nomura, T. Takahashi, Y. Yokomizo, Ferroelectric properties in the system Pb(Zn$_{1/3}$Nb$_{2/3}$)O$_3$-PbTiO$_3$, J. Phys. Soc. Japan. 27 (1969) 262. doi:10.1143/JPSJ.27.262.

[5]   B. Fang, H. Xu, H. Luo, Growth and properties of Pb[(Zn$_{1/3}$Nb$_{2/3}$)$_{0.91}$Ti$_{0.09}$] O$_3$ single crystals directly from melt, J. Cryst. Growth. 310 (2008) 2871–2877. doi:10.1016/j.jcrysgro.2008.01.026.

[6]   A. Dąbkowski, H.A. Dąbkowska, J.E. Greedan, W. Ren, B.K. Mukherjee, Growth and properties of single crystals of relaxor PZN-PT materials obtained from high-temperature solution, J. Cryst. Growth. 265 (2004) 204–213. doi:10.1016/j.jcrysgro.2004.01.057.

[7]   T. Kobayashi, S. Saitoh, K. Harada, S. Shimanuki, Y. Yamashita, Growth of large and homogeneous PZN-PT single crystals for medical ultrasonic array transducers, IEEE Int. Symp. Appl. Ferroelectr. (1998) 235–238. doi:10.1109/isaf.1998.786677.

[8]   S. Ganesamoorthy, G. Singh, I. Bhaumik, A.K. Karnal, V.S. Tiwari, K. Kitamura, V.K. Wadhawan, Growth of relaxor ferroelectric single crystals PbZn$_{1/3}$Nb$_{2/3}$O$_3$ (PZN) by high temperature solution growth, Ferroelectrics. 326 (2005) 19–23. doi:10.1080/00150190500318131.

[9]   Y. Huang, Y. Xia, Di.H. Lin, K. Yao, L.C. Lim, Large stroke high fidelity PZN-PT single-crystal "stake" Actuator, IEEE Trans. Ultrason. Ferroelectr. Freq. Control. 64 (2017) 1617–1624. doi:10.1109/TUFFC.2017.2735800.

[10]  L.M. Ewart, E.A. McLaughlin, H.C. Robinson, A. Amin, J.J. Stace, Mechanical and electromechanical properties of PMN-PT single crystals for naval sonar transducers, IEEE Int. Symp. Appl. Ferroelectr. 54 (2007) 553–556. doi:10.1109/ISAF.2007.4393327.

[11]  C. Manikandan, E. Varadarajan, P. Vijayakumar, R. Ramesh, V. Roy, R.M. Sarguna, E.P. Amaladass, S. Ganesamoorthy, T.K. Vinodkumar, M.N. Unni, C.S.N. Venkataraman, V. Natarajan, S.M. Babu, Realization of high performance PZN-PT single crystal based piezoelectric flexural mode hydrophone for underwater sensor

applications, Mater. Res. Express. 10 (2023) 066303. doi:10.1088/2053-1591/acdfbd.

[12] B. Srimathy, R. Jayavel, I. Bhaumik, S. Ganesamoorthy, A.K. Karnal, P.K. Gupta, J. Kumar, Role of dopant induced defects on the properties of Nd and Cr doped PZNT single crystals, Mater. Sci. Eng. B. 185 (2014) 60–66. doi:10.1016/j.mseb.2014.01.018.

[13] S. Zhang, F. Li, F. Yu, X. Jiang, H.Y. Lee, J. Luo, T.R. Shrout, Recent developments in piezoelectric crystals, J. Korean Ceram. Soc. 55 (2018) 419–439. doi:10.4191/kcers.2018.55.5.12.

[14] M.L. Mulvihill, S.E. Park, G. Risch, Z. Li, K. Uchino, T.R. Shrout, The role of processing variables in the flux growth of lead zinc niobate-lead titanate relaxor ferroelectric single crystals, Japanese J. Appl. Physics, 35 (1996) 3984–3990. doi:10.1143/jjap.35.3984.

[15] Z.G. Ye, Handbook of Advanced Dielectric, Piezoelectric and Ferroelectric Materials: Synthesis, Properties and Applications, Woodhead Publishing Limited, 2008, Cambridge, United Kingdom. doi:10.1533/9781845694005.

[16] Y. Tyagur, I. Tyagur, A. Kopal, L. Burianova, P. Hana, Dielectric and piezoelectric properties of $Sn_2P_2S_6$ single crystals, Ferroelectrics. 320 (2005) 35–42. doi:10.1080/00150190590966766.

[17] J. Luo, S. Zhang, Advances in the growth and characterization of relaxor-PT-based ferroelectric single crystals, Crystals. 4 (2014) 306–330. doi:10.3390/cryst4030306.

[18] S. Zhang, F. Li, J. Luo, R. Sahul, T. Shrout, Relaxor-$PbTiO_3$ single crystals for various applications, IEEE Trans. Ultrason. Ferroelectr. Freq. Control. 60 (2013) 1572–1580. doi:10.1109/TUFFC.2013.2737.

[19] L.C. Lim, K.K. Rajan, High-homogeneity high-performance flux-grown $Pb(Zn_{1/3}Nb_{2/3})O_3$-(6-7)%$PbTiO_3$ single crystals, J. Cryst. Growth. 271 (2004) 435–444. doi:10.1016/j.jcrysgro.2004.07.081.

[20] S. Zhang, F. Li, High performance ferroelectric relaxor-$PbTiO_3$ single crystals: Status and perspective, J. Appl. Phys. 111 (2012) 031301. doi:10.1063/1.3679521.

[21] K. Maria Kuruvila, D. Dhayanithi, S. Manivannan, N. V. Giridharan, P. Vijayakumar,

C. Manikandan, R.M. Sarguna, E. Prabu Amaladass, S. Ganesamoorthy, E. Varadarajan, V. Natarajan, A study on the electrical properties of flux grown 0.91PZN-0.09PT single crystals for high-performance piezoelectric and pyroelectric device applications, J. Cryst. Growth. 598 (2022) 126875. doi:10.1016/j.jcrysgro.2022.126875.

[22] C. Zou, Y. Li, S. Hou, Z. Liu, H. Tang, S. Chen, J. Peng, Development of cardiac phased array with large-size PZN-5.5 %PT single crystals, IEEE Trans. Ultrason. Ferroelectr. Freq. Control. 69 (2022) 744–750. doi:10.1109/TUFFC.2021.3120774.

[23] G. Singh, I. Bhaumik, S. Ganesamoorthy, A.K. Karnal, V.S. Tiwari, Domain structure and birefringence studies on a 0.91Pb(Zn$_{1/3}$Nb$_{2/3}$)O$_3$ -0.09PbTiO$_3$ single crystal, Cryst. Res. Technol. 42 (2007) 378–383. doi:10.1002/crat.200610831.

[24] I.E. Myl'nikova, V.A. Bokov A.V. Shubnikov, N.N. Sheftal (Eds.), Growth of Crystals, vol. 3, Consultants Bureau, New York (1962), p. 309

[25] R. Bertram, G. Reck, R. Uecker, Growth and correlation between composition and structure of (1-$x$)Pb(Zn$_{1/3}$Nb$_{2/3}$)O$_3$-$x$PbTiO$_3$ crystals near the morphotropic phase boundary, J. Cryst. Growth. 253 (2003) 212–220. doi:10.1016/S0022-0248(03)00972-2.

[26] F.J. Kumar, L.C. Lim, C. Chilong, M.J. Tan, Morphological aspects of flux grown 0.91PZN-0.09PT crystals, J. Cryst. Growth. 216 (2000) 311–316. doi:10.1016/S0022-0248(00)00446-2.

[27] Y.H. Li, Y.Q. Wang, M. Zhou, C.P. Xu, J.A. Valdez, K.E. Sickafus, Light ion irradiation effects on stuffed Lu$_2$(Ti$_{2-x}$Lu$_x$)O$_{7-x/2}$ ($x$ = 0, 0.4 and 0.67) structures, Nucl. Instruments Methods Phys. Res. Sect. B Beam Interact. with Mater. Atoms. 269 (2011) 2001–2005. doi:10.1016/j.nimb.2011.05.036.

[28] M. Matsushita, Y. Tachi, K. Echizenya, Growth of 3-in single crystals of piezoelectric Pb[(Zn$_{1/3}$Nb$_{2/3}$)$_{0.91}$Ti$_{0.09}$]O$_3$ by the supported solution Bridgman method, J. Cryst. Growth. 237–239 (2002) 853–857. doi:10.1016/S0022-0248(01)02052-8.

[29] B. Fang, H. Luo, H. Xu, T. He, Z. Yin, Growth of Pb[(Zn$_{1/3}$Nb$_{2/3}$)$_{0.91}$Ti$_{0.09}$]O$_3$ single crystals directly from melt, Jpn. J. Appl. Phys. 40 (2001) L1377.

doi:10.1143/JJAP.40.L1377.

[30] T. Kobayashi, S. Shimanuki, S. Saitoh, Y. Yamashita, Improved growth of large lead zinc niobate titanate piezoelectric single crystals for medical ultrasonic transducers, Japanese J. Appl. Physics, 36 (1997) 6035–6038. doi:10.1143/jjap.36.6035.

[31] J. Xu, X. Wu, J. Tong, M. Shi, G. Qian, Two-step Bridgman growth of $0.91Pb(Zn_{1/3}Nb_{2/3})O_3$-$0.09PbTiO_3$ single crystals, J. Cryst. Growth. 280 (2005) 107–112. doi:10.1016/j.jcrysgro.2005.02.067.

[32] K. Uchino, Advanced Piezoelectric Materials: Science and Technology, Woodhead Publishing Limited, 2010, Cambridge, United Kingdom. doi:10.1533/9781845699758.